\documentclass[10pt]{article}
\usepackage{epsfig}
\def\beq{\begin{equation}}
\def\eeq#1{\label{#1}\end{equation}}
\def\eeqn{\end{equation}}

\def\beqa{\begin{eqnarray}}
\def\eeqa#1{\label{#1}\end{eqnarray}}
\def\eeqan{\end{eqnarray}}

\let\bar=\overbar

\def\W{{\cal W}}

\def\Dslash{\not{\hbox{\kern-4pt $D$}}}
\def\dslash{\not{\hbox{\kern-2pt $\del$}}}

\def\msb{{\bar{\ssstyle M \kern -1pt S}}}

\def\Title#1{\begin{center} {\Large {\bf #1} } \end{center}}
\usepackage{lineno}
\begin{document}

\Title{Measurement of the $t\bar{t}$ spin correlations and top quark polarization in dileptonic channel with the CMS detector}

\bigskip\bigskip


\begin{raggedright}  

{\it Ajeeta Khatiwada on behalf of the CMS Collaboration\index{Khatiwada, A.}\\
Department of Physics\\
Purdue University\\
West Lafayette, IN\\
Email: ajeeta.khatiwada@cern.ch}
\bigskip\bigskip
\end{raggedright}

Talk presented at the APS Division of Particles and Fields Meeting (DPF 2017), July 31-August 4, 2017, Fermilab. C170731

\begin{abstract}
The degree of top polarization and strength of $t\bar{t}$ correlation are dependent on production dynamics, decay mechanism, and choice of the observables. At the LHC, measurement of the top polarization and spin correlations in $t\bar{t}$ production is possible through various observables related to the angular distribution of decay leptons. A measurement of differential distribution provides a precision test of the standard model of particle physics and probes for deviations, which could be a sign of new physics. In particular, the phase space for the super-symmetric partner of the top quark can be constrained. Results from the Compact Muon Solenoid (CMS) collaboration for top quark polarization and spin correlation in the dileptonic channel are reviewed briefly in this proceeding. The measurements are obtained using 19.5 fb$^{-1}$ of data collected in pp collisions at the center-of-mass energy of 8 TeV.
\end{abstract}

\section{Introduction}
The top quark is the heaviest known fundamental particle and has a very short lifetime. With a mass close to the electroweak scale, it is expected to play a crucial role in electroweak symmetry breaking. Moreover, at the order of 1 GeV, its decay width ($\Gamma_{t}$) is larger than the QCD hadronization scale $(\lambda_{QCD}\sim 0.1$ GeV) and much larger compared to the spin decorrelation scale ($\lambda_{QCD}^2/m_t\sim 0.1$ MeV)\cite{Mahlon1}. Hence, the top quark decays before hadronization occurs and its spin information is carried by the angular distribution of the decay products. \\
\\
At the LHC, the top quark is produced predominantly in pair via gluon fusion ($gg\rightarrow t\bar{t}$). The top quarks produced as a $t\bar{t}$ pair are mostly unpolarized due to the parity conserving nature of the strong interaction. A higher order electroweak correction to the QCD gives a small net polarization to the top quarks and the spin of the top is correlated with the spin of the anti-top. At low invariant mass, the top pair production is dominated by like helicity pairs, meanwhile, the unlike helicity pairs dominate the production at the higher invariant mass regime\cite{Mahlon2}. For many beyond the standard model scenarios, the top quarks may be produced polarized. The presence of new physics (NP) may also alter the Standard Model (SM) $t\bar{t}$ spin correlation. Therefore, the precision study of spin correlation in $t\bar{t}$ pair and the top polarization serves to validate the SM prediction and to probe for deviations, which indicate signs of new physics. In this report, the results from spin correlation and polarization measurements at 8 TeV in the dileptonic channel\cite{8TeVspincorr} using the CMS\cite{cmsdet} detector are presented, and improvements being made in an ongoing analysis at 13 TeV are discussed.\\
\section{Observables}
In SM $t\bar{t}$ production, the spin information of top and anti-top are carried by their respective decay products. Therefore, the $t\bar{t}$ correlation leads to a correlation between the decay products. The differential cross section for $t\bar{t}$ production can be expressed as, 
\begin{equation}
\frac{1}{\sigma}\frac{d\sigma}{d\cos\theta_{l^+} d\cos\theta_{l^-}} = \frac{1}{4}(1+B_1\cos\theta_{l^+} + B_2\cos\theta_{l^-} - C\cos\theta_{l^+}\cos\theta_{l^-})\cite{Mahlon2}.
\end{equation}
Here, $B_1$ ($B_2$) is the polarization in quantization axis $\hat{a}$ ($\hat{b}$) and $C$ is the spin correlation. $\theta_{l^+}$($\theta_{l^-}$) is the angle made by the decay product with respect to the quantization axes, $\hat{a}$ ($\hat{b}$), in the rest frame of the parent top (anti-top).  $C$ can further be expressed in terms of spin analyzing power, $\kappa$, as, 
\begin{equation}
C(\hat{a},\hat{b}) = \kappa^2 \frac{\sigma(\uparrow\uparrow)+\sigma(\downarrow\downarrow)-\sigma(\uparrow\downarrow)-\sigma(\downarrow\uparrow)}{\sigma(\uparrow\uparrow)+\sigma(\downarrow\downarrow)+\sigma(\uparrow\downarrow)+\sigma(\downarrow\uparrow)}\cite{Mahlon2}.
\end{equation} 
In top decays, $\kappa$ is +1.0 for charged lepton and down quark, while $\kappa$ is -0.41 for bottom quark and -0.31 for up type quark and neutrino. This makes the dileptonic mode ($t\bar{t}\rightarrow\W^+b\W^-b\rightarrow b l^+\nu_l \bar{b} l^- \bar{\nu_l}$) a better probe of the top spin than other decay modes with jets in the final state\cite{Mahlon3}. Although the dileptonic channel has a lower branching ratio when compared to the all-hadronic channel (both $W$ bosons from top and anti-top decay to quark anti-quark pair) and the semi-leptonic channel (one $W$ boson decays to quark/anti-quark and the other decays to leptons), the signal to background ratio is better. In the above equation, $\sigma(\uparrow\downarrow)$ and $\sigma(\downarrow\uparrow)$ refer to the cross section of events with anti-correlated top spins and $\sigma(\uparrow\uparrow)$ and $\sigma(\downarrow\downarrow)$ refer to the cross section of events with correlated spins. \\
\\
In the 8 TeV analysis by CMS\cite{8TeVspincorr}, the variable $\cos\theta_l^*$, as measured in helicity basis (defined by the direction of the parent top quark's momentum), was studied as a probe to the polarization. The asymmetry related to $\cos\theta_{l^\pm}^*$, $A_{P^\pm}$, is calculated as, 
\begin{equation}
A_{P^\pm} = \frac{N[\cos\theta_{l^\pm}^*>0]-N[\cos\theta_{l^\pm}^*<0]}{N[\cos\theta_{l^\pm}^*>0]+N[\cos\theta_{l^\pm}^*<0]}.
\end{equation}
$A_P$ is related to the SM polarization, $P=2A_P=(A_{P^+}+A_{P^-})$, and the polarization associated to the maximally CP-violating process, $P^{CPV}=2A_P^{CPV}=(A_{P^+}-A_{P^-})$.\\
\\
The product of the polarization variable for lepton and anti-lepton probes the spin correlation and its asymmetry, $A_{c_1c_2}$, where $c_1 = \cos\theta^*_{l^+}$ and $c_2 = \cos\theta^*_{l^-}$, gives a measure of the amount of spin correlation,
\begin{equation}
A_{c_1c_2} = \frac{N[c_1c_2>0]-N[c_1c_2<0]}{N[c_1c_2>0]+N[c_1c_2<0]}.
\end{equation}
The amount of spin correlation is given by $C_{hel}=-4A_{c_1c_2}$.\\
\\
Another spin correlation observable $\cos\phi$, where $\phi$ is the angle between the leptons in their respective parents' rest frame, was also studied as a probe to spin correlation. The asymmetry related to $\cos\phi$ is, 
\begin{equation}
A_{\cos\phi} = \frac{N[\cos\phi>0]-N[\cos\phi<0]}{N[\cos\phi>0]+N[\cos\phi<0]}.
\end{equation}
The asymmetry related to $\cos\phi$ gives the measurement of spin correlation coefficient $D$ by the relation $D=-2A_{cos\phi}$.\\
\\
The opening angle between two leptons in the lab frame, $\Delta\phi_{l\bar{l}}$, which does not require full event reconstruction, was also studied. The asymmetry for $\Delta\phi_{l\bar{l}}$ variable is, 
\begin{equation}
A_{\Delta\phi} = \frac{N[|\Delta\phi_{l\bar{l}}|>\pi/2]-N[|\Delta\phi_{l\bar{l}}|<\pi/2]}{N[|\Delta\phi_{l\bar{l}}|>\pi/2]+N[|\Delta\phi_{l\bar{l}}|<\pi/2]}.
\end{equation}

\section{Event Selection}
Events were selected using double lepton triggers with transverse momentum ($p_T$) greater than 17 (8) GeV for the leading (sub-leading) lepton. The offline selection required exactly two oppositely charged isolated leptons with transverse momentum $(p_T^l)>20$ GeV and $|\eta^l|<2.4$. Isolation was measured by summing the scalar $p_T$ of all the particle flow objects that are not associated with the lepton and in the cone of radius $\Delta R = \sqrt{(\Delta\eta)^2+(\Delta\phi)^2}$ = 0.3.  It was required to be less than $0.15~p_T^l)$ or $5$ GeV to ensure that the leptons are not part of the jet. Furthermore, contributions from pileup were subtracted from the isolation prior to applying the cut. Background from low mass Drell-Yan processes were rejected by requiring the lepton invariant mass to be greater than 20 GeV. In the case of same flavor lepton final state, lepton invariant mass within 15 GeV of the Z boson mass were also removed to suppress backgrounds from Z boson decay. The magnitude of the missing transverse momentum ($E_T^{miss}$), commonly referred to as the missing transverse energy, was also required to be greater than 40 GeV in the same-flavor lepton events to further suppress Drell-Yan background. At least two anti-$k_T$ jets\cite{antiktjetalgo} with a distance parameter of $\Delta R=0.5$, having $p_T^{j}$ greater than 30 GeV and $|\eta^{j}|$ less than 2.4, were required. At least one of the jets is required to be consistent with originating from b-quarks as per the medium working point definition of the CSV b-tagging algorithm\cite{btaggingalgo}. Since, all but one of the observables in the study require reconstruction of the $t\bar t$ system, analytical solutions for the two neutrinos' four momenta were obtained from the constraints on the invariant masses of the top quark and W bosons, as well as from the constraints on the missing transverse momentum vector $\vec{p}_T^{miss}$ in the transverse direction\cite{analyticalsolcms},\cite{dalitz},\cite{betchart}.\\
\section{Results}
\begin{figure}[htb]
\begin{center}
\includegraphics[height=2.2in]{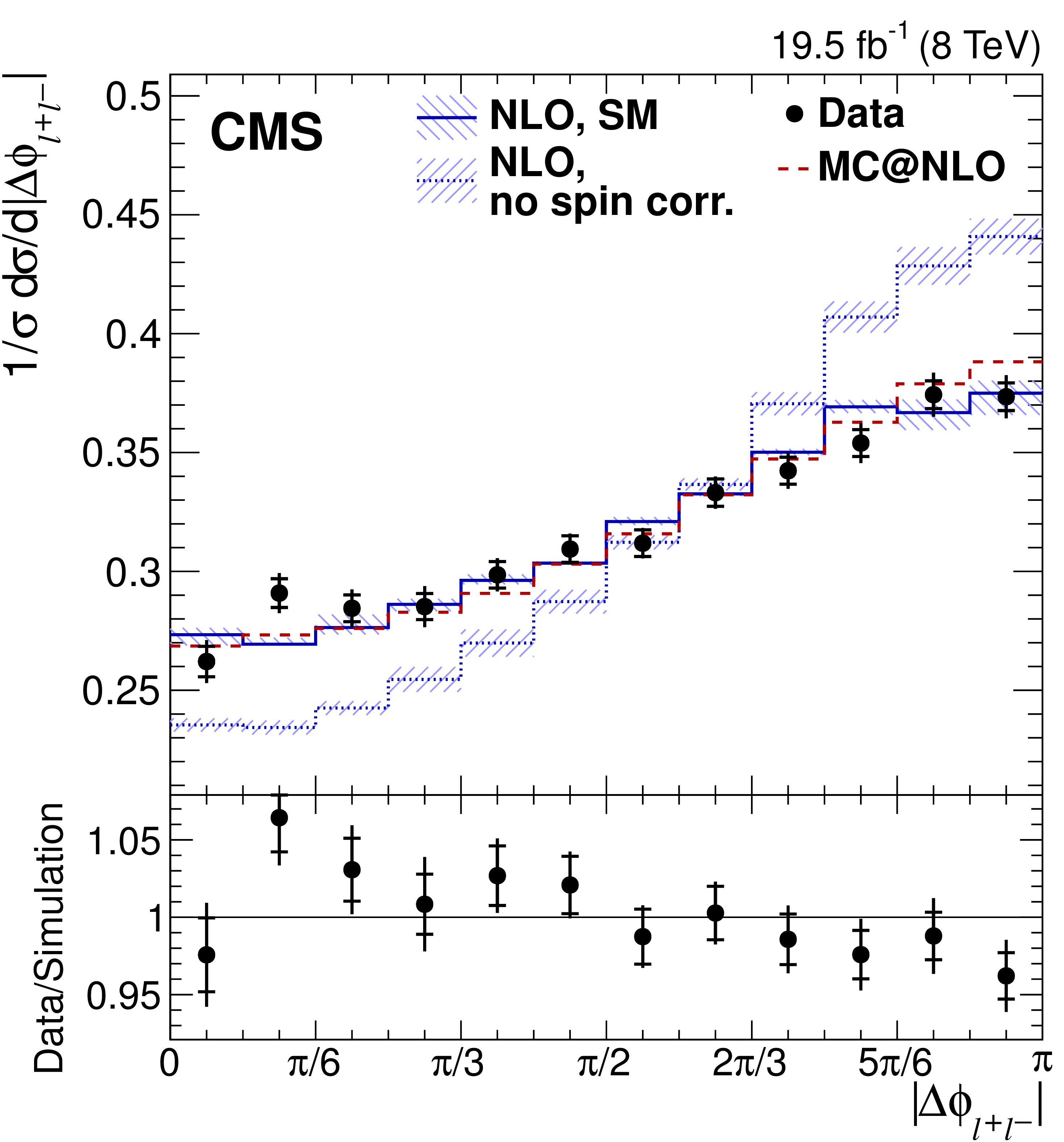}
\includegraphics[height=2.2in]{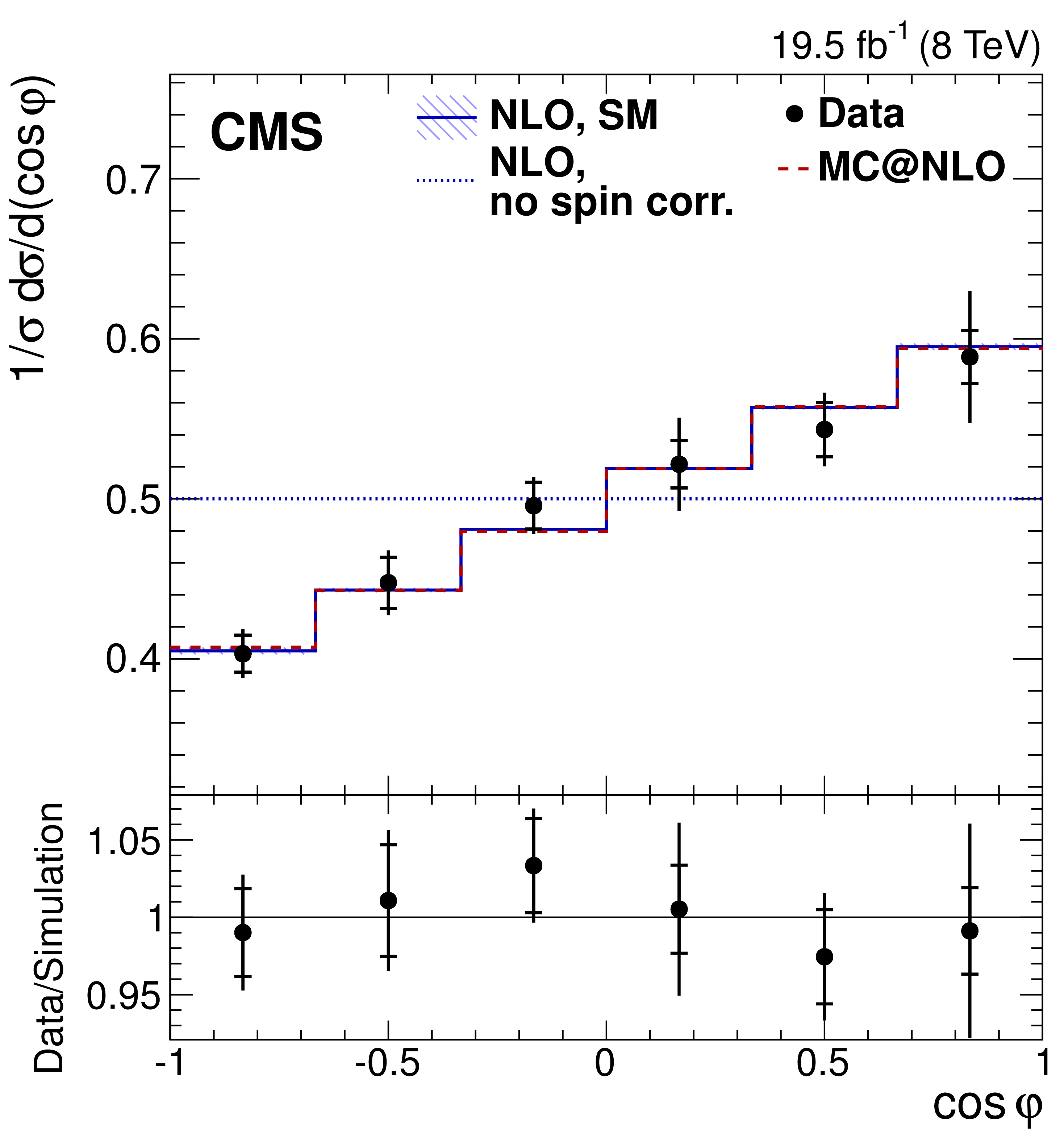}\\
\includegraphics[height=2.2in]{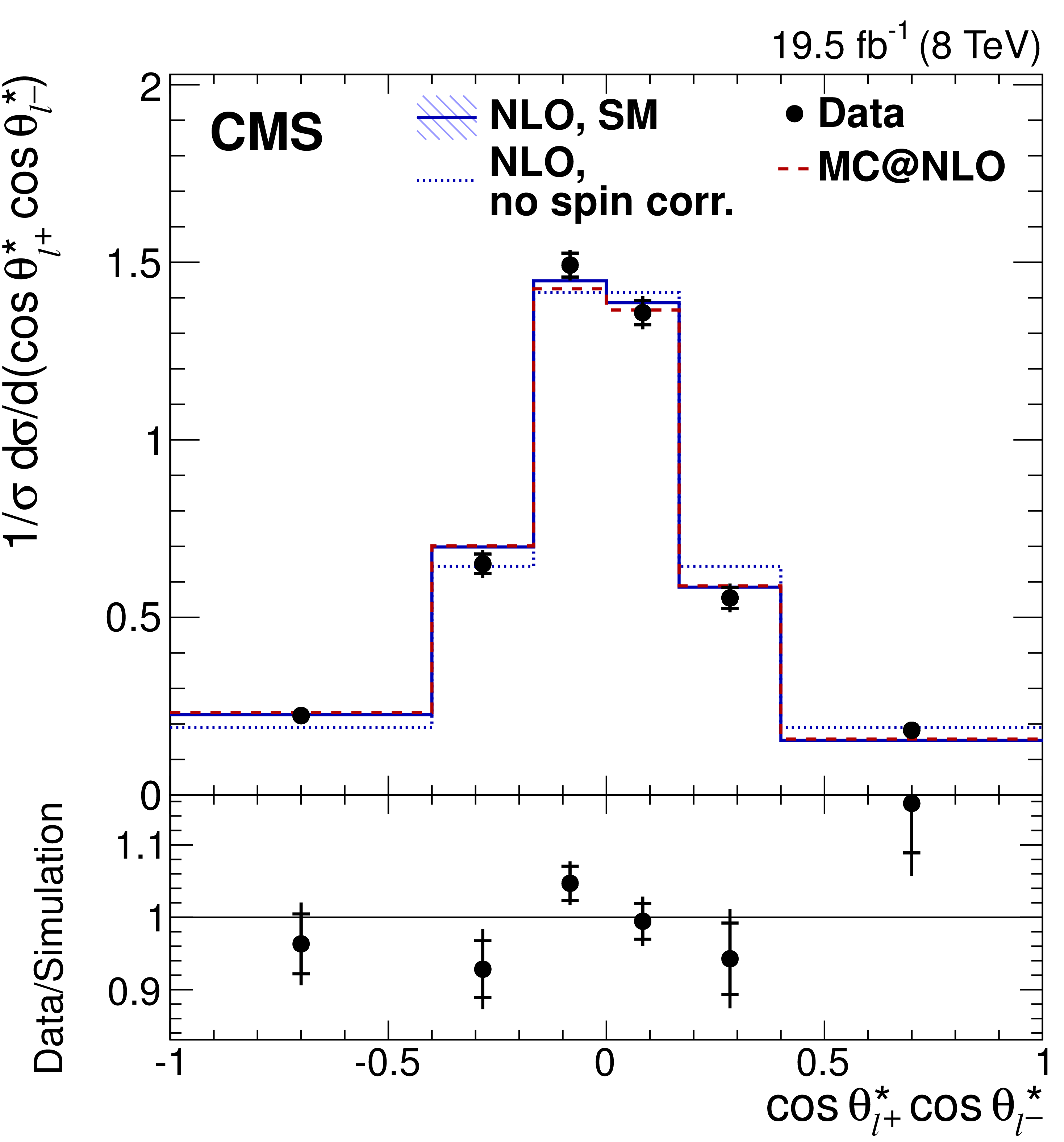}
\includegraphics[height=2.2in]{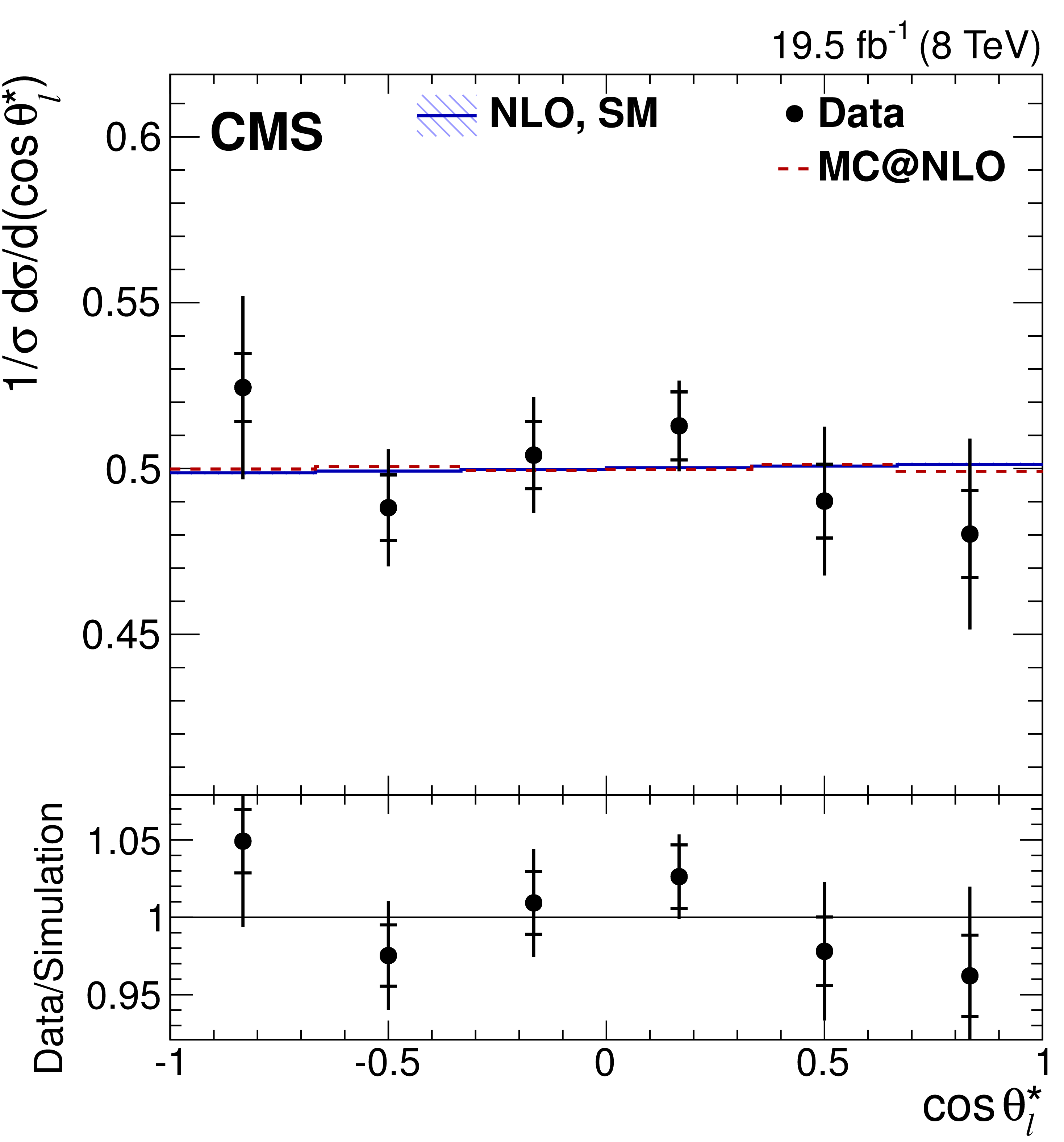}
\caption{Differential cross section as a function of $\Delta\phi_{l^+ l^-}$, $cos\phi$, $\cos\theta_{l}^*$, $\cos\theta_{l^+}^*\cos\theta_{l^-}^*$ (clockwise from top-left to bottom-right) compared to parton-level predictions from MC@NLO (dashed red lines), theoretical prediction at next to leading order with SM spin correlation (blue solid line), and similar theoretical prediction without the spin correlation (dotted blue). The inner (outer) bars in the data points represent statistical (total) uncertainties in the measurement. The hatched area in the theoretical calculation represent scale uncertainty from the systematic variation of renormalization scale ($\mu_R$) and factorization scale ($\mu_F$) by a factor of two\cite{8TeVspincorr}.}
\label{fig:1Ddist}
\end{center}
\end{figure}

\begin{figure}[htb]
\begin{center}
\includegraphics[height=1.8in]{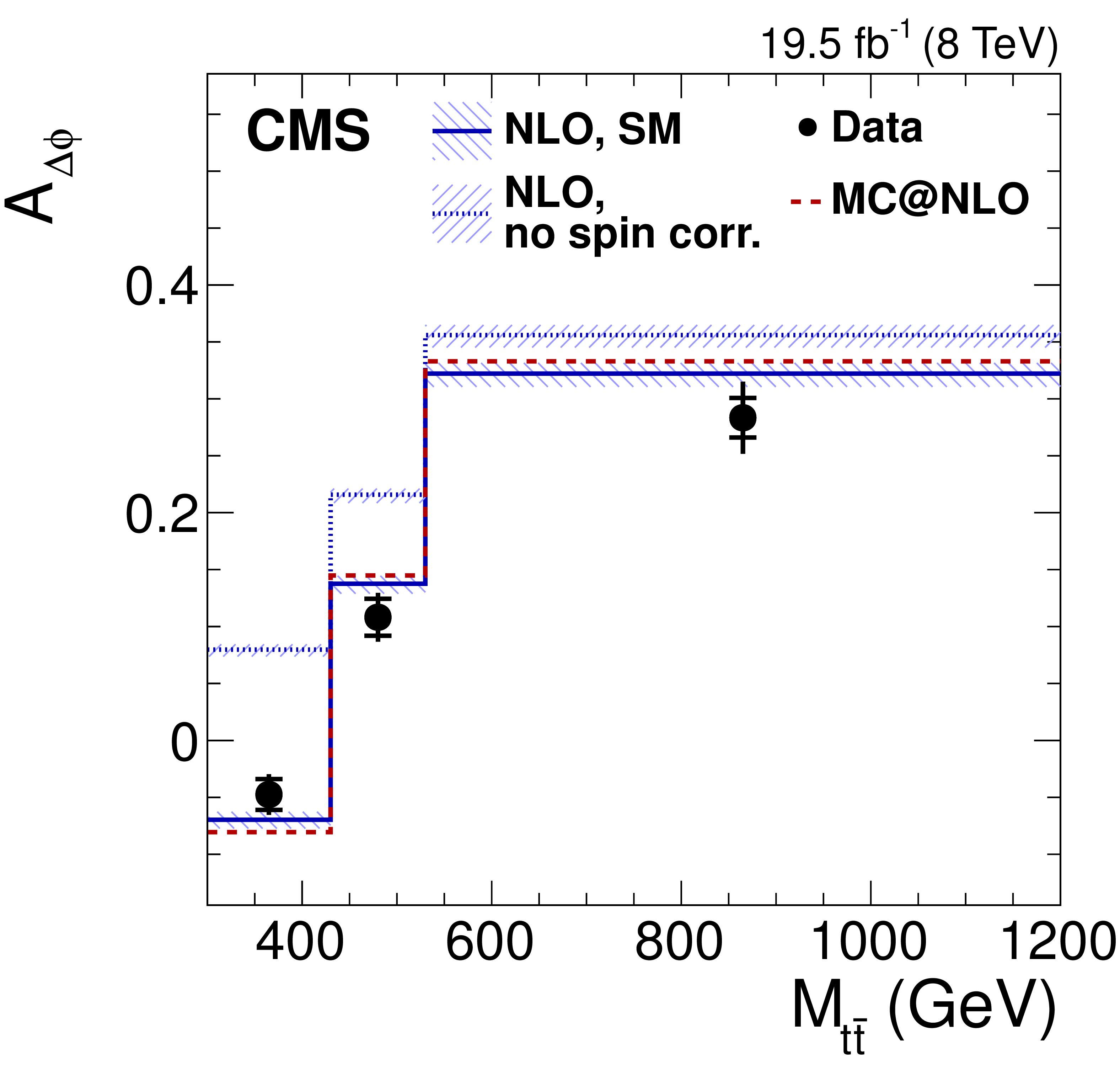}
\includegraphics[height=1.8in]{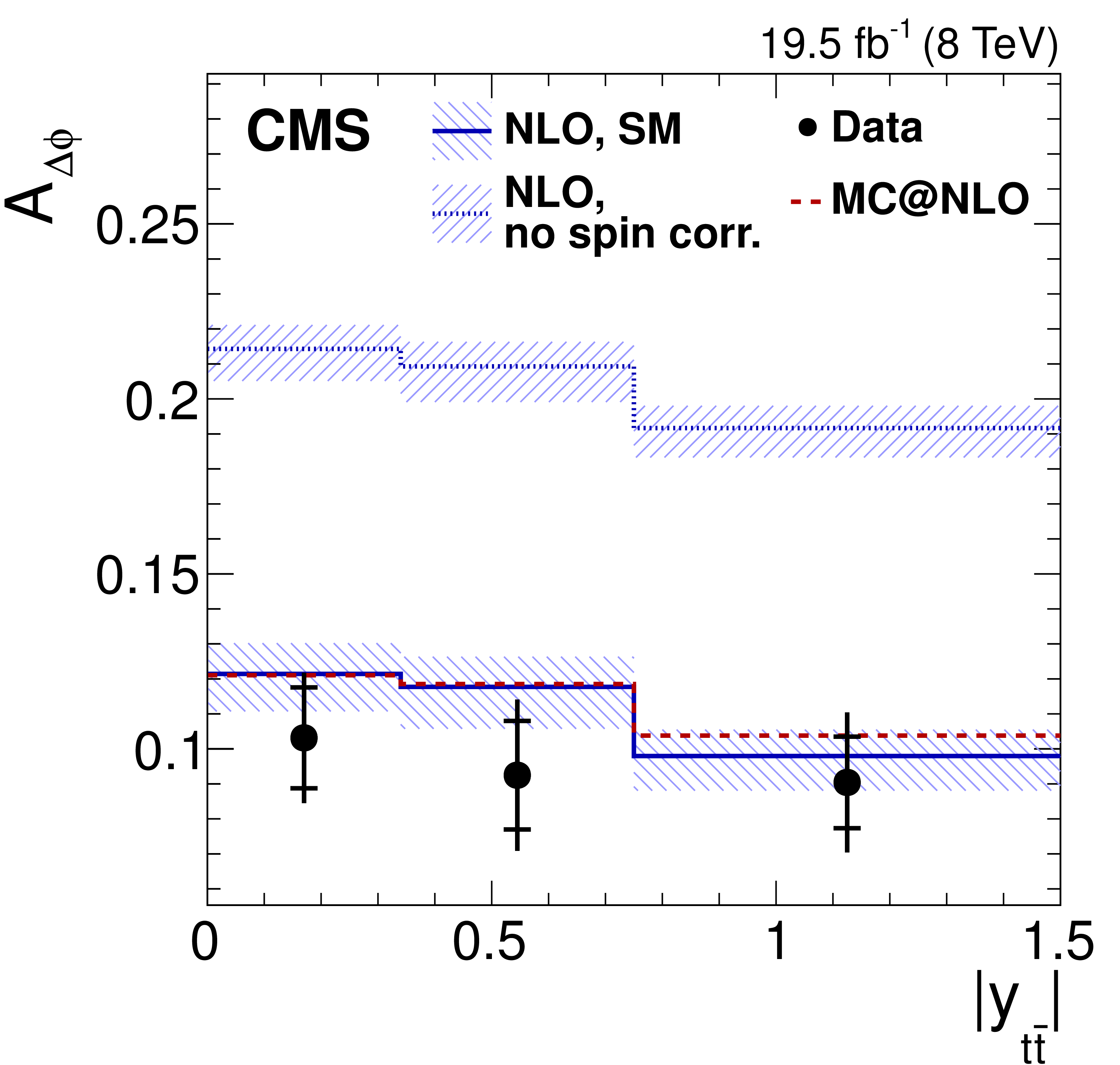}
\includegraphics[height=1.8in]{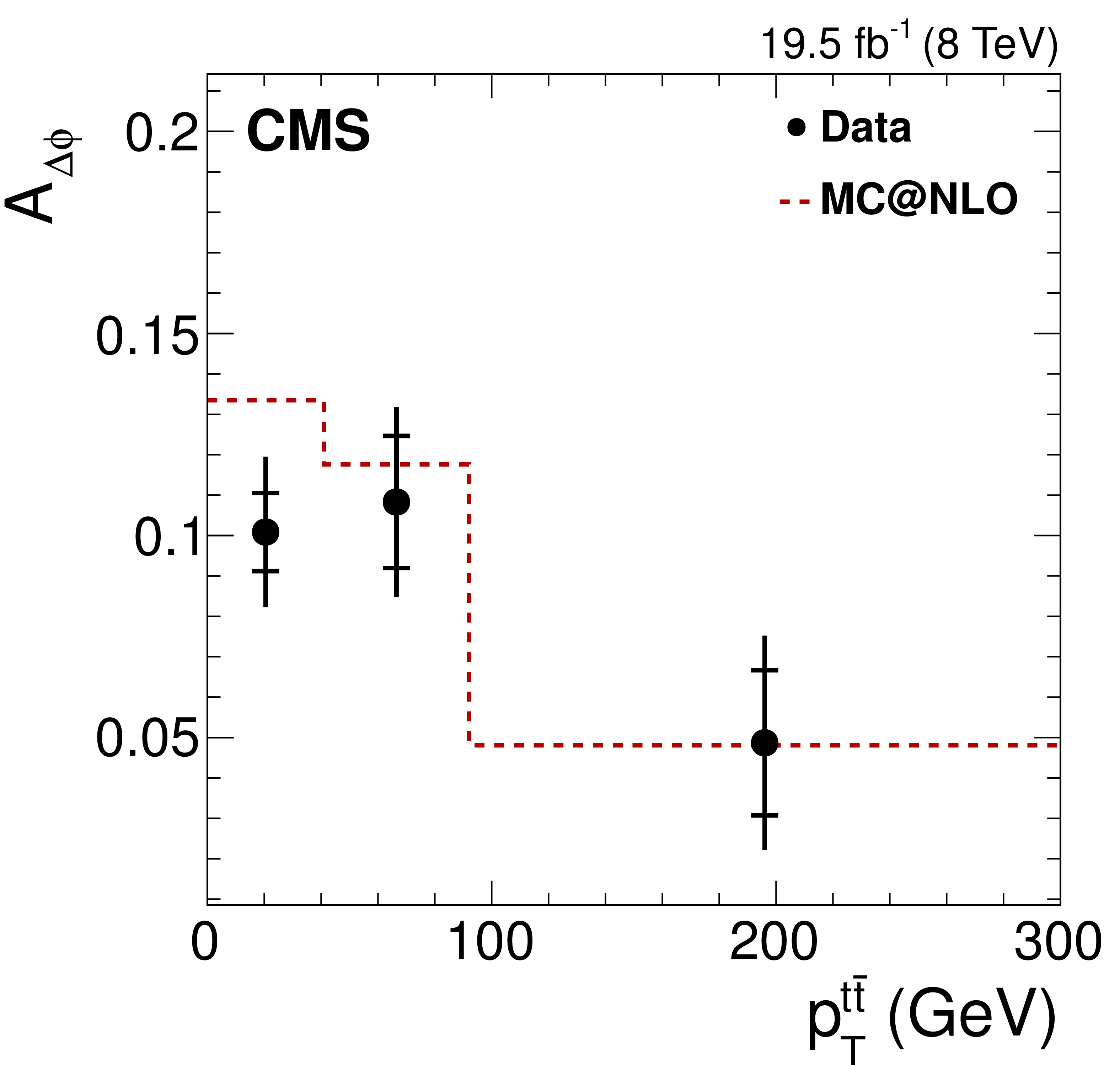}\\
\includegraphics[height=1.8in]{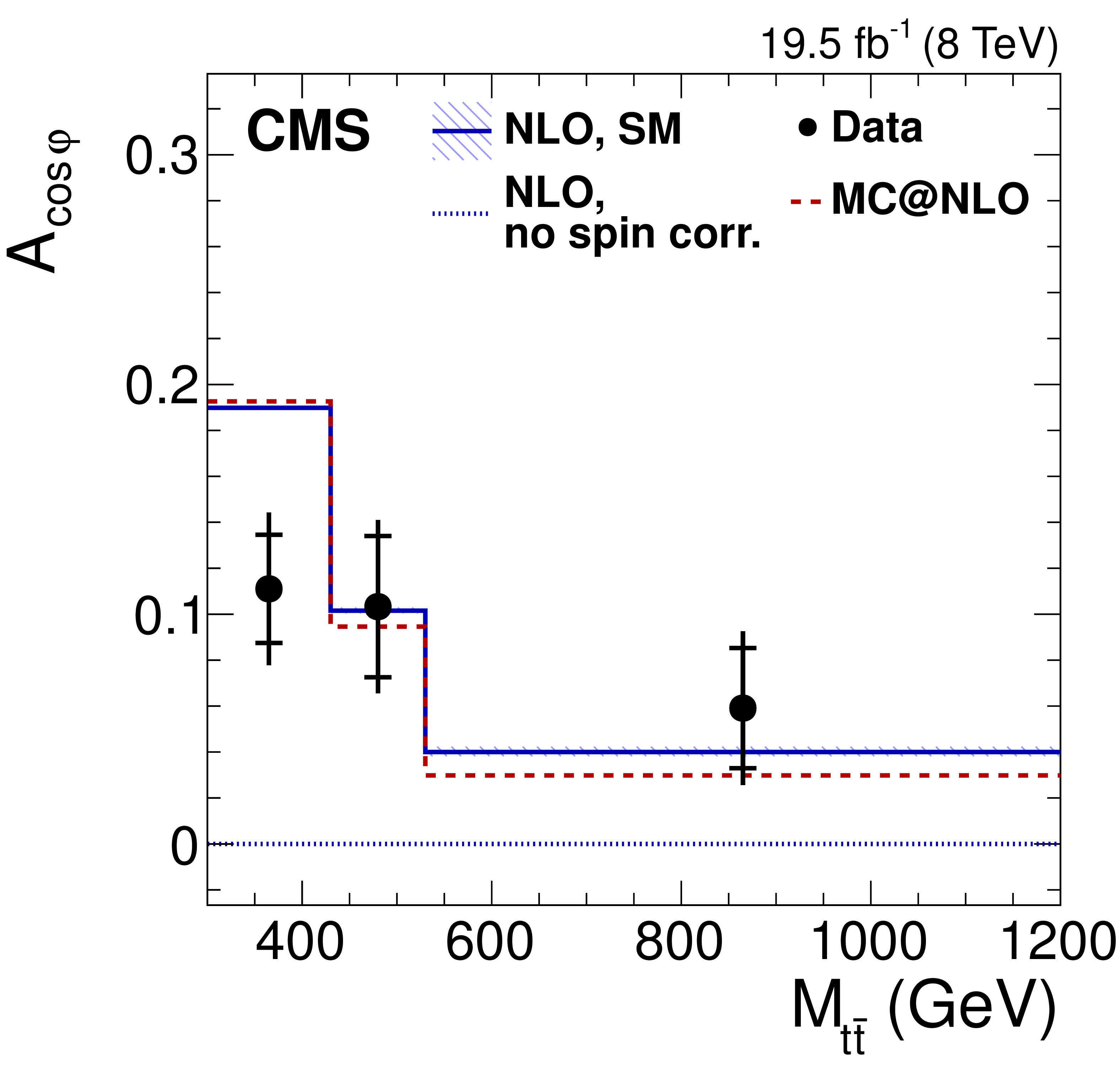}
\includegraphics[height=1.8in]{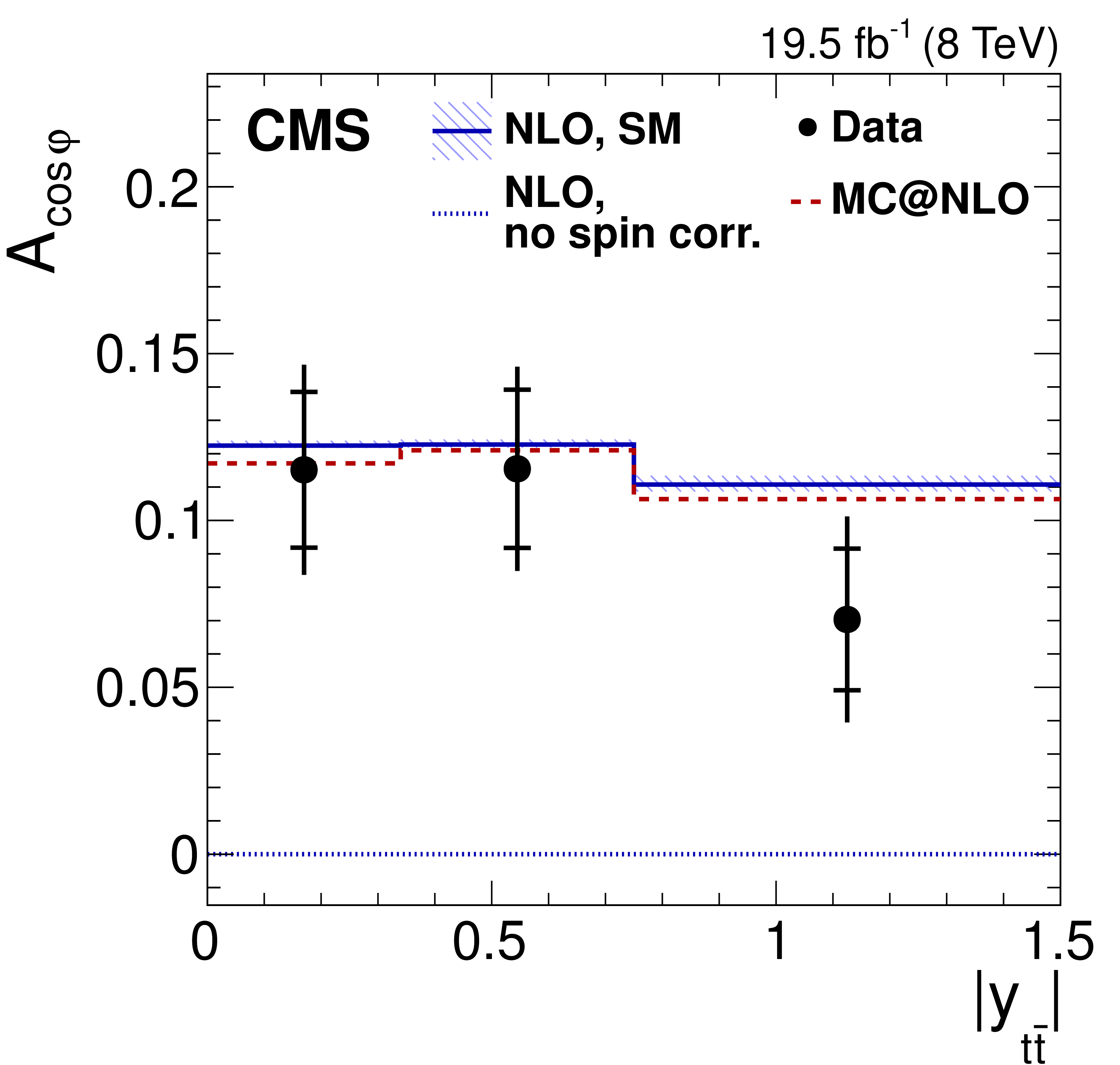}
\includegraphics[height=1.8in]{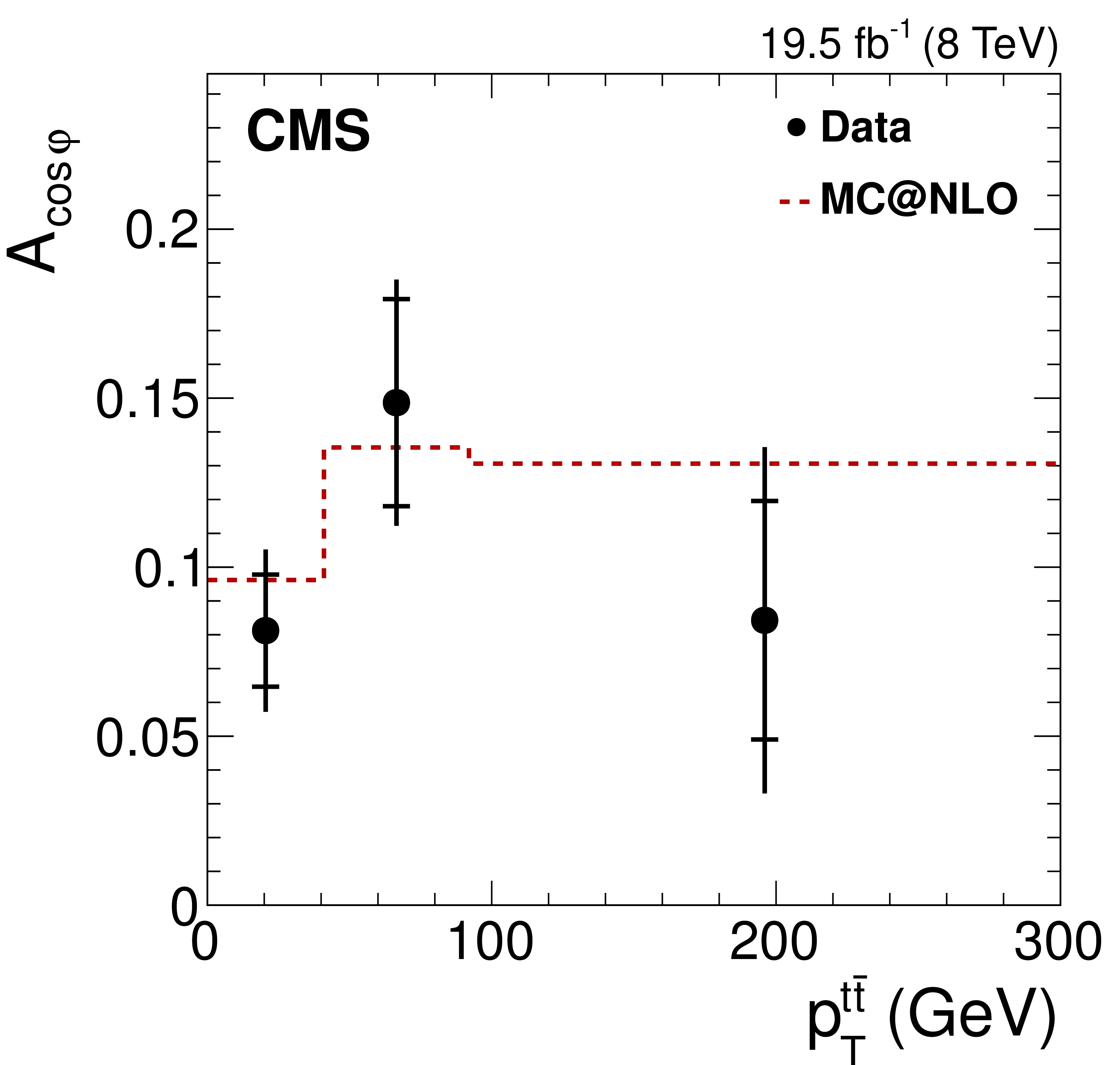}\\
\includegraphics[height=1.8in]{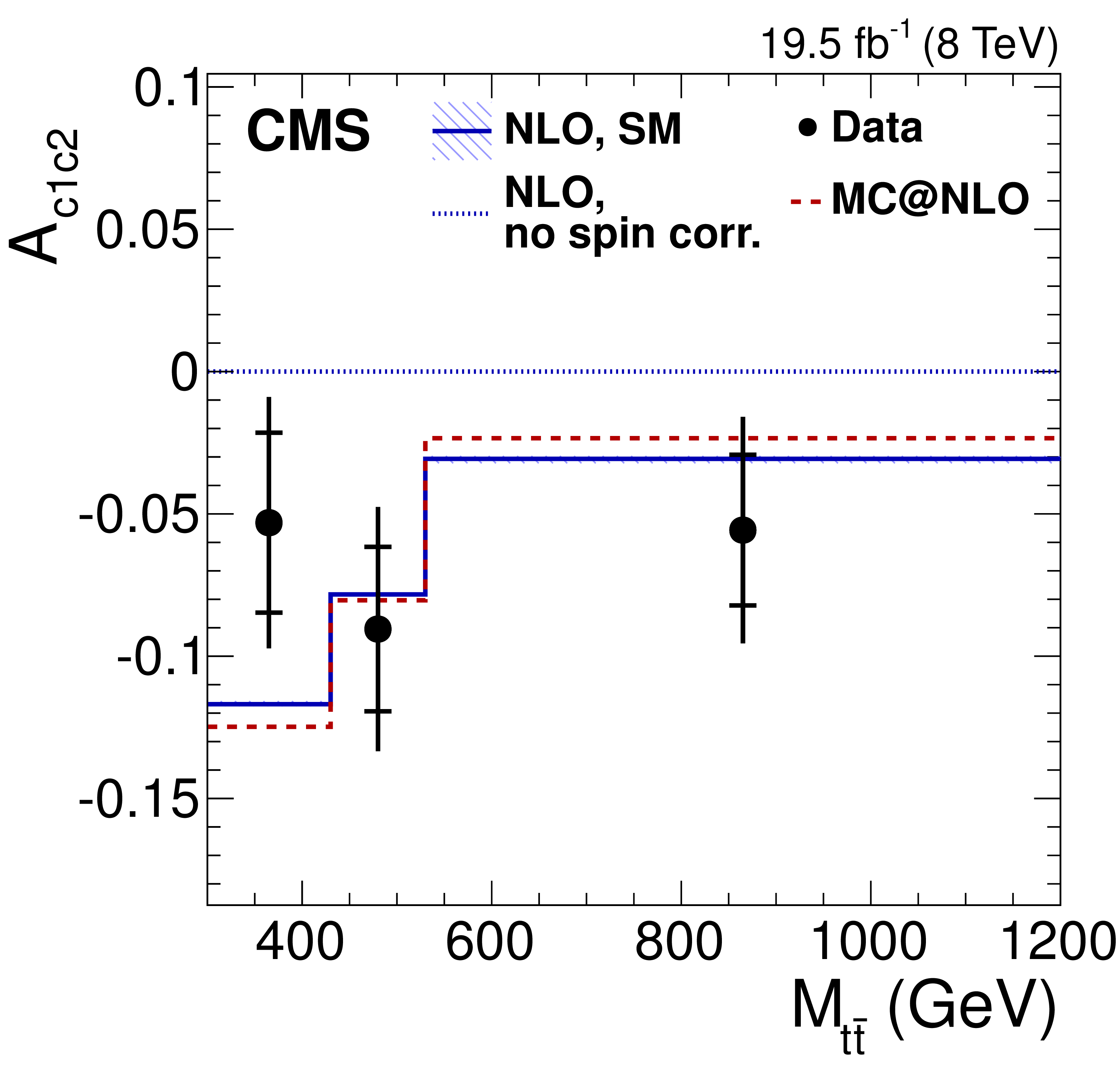}
\includegraphics[height=1.8in]{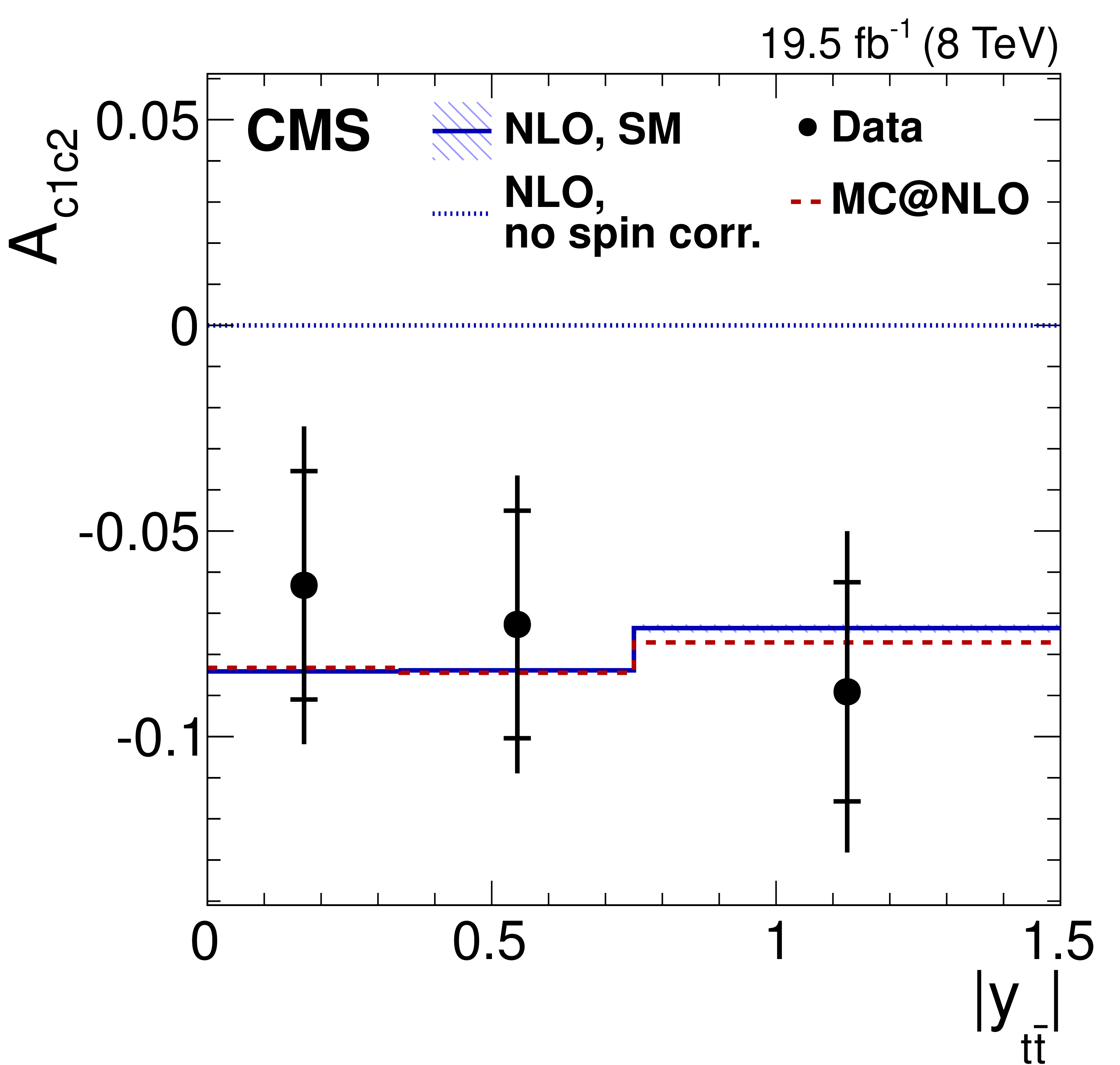}
\includegraphics[height=1.8in]{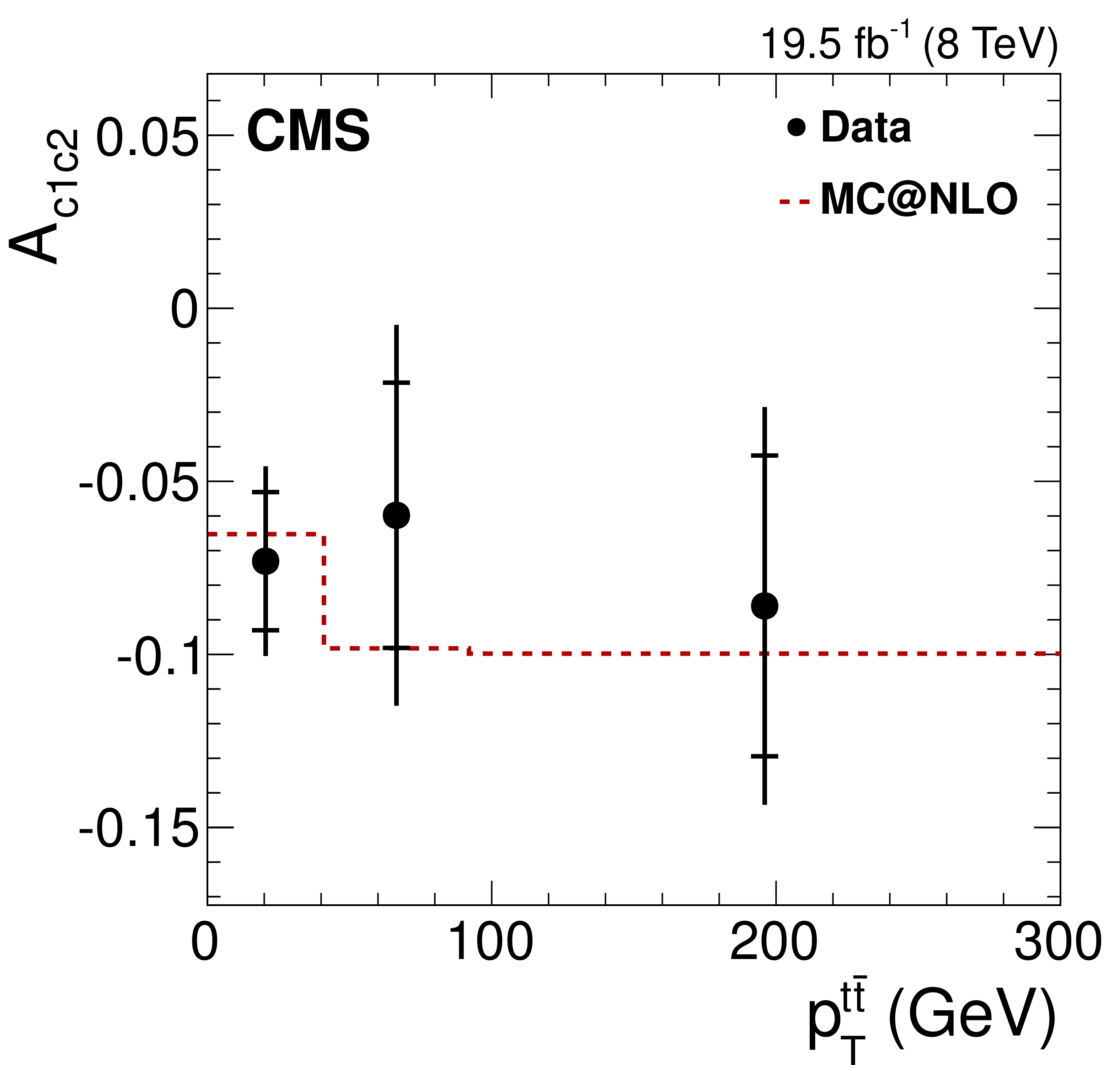}\\
\includegraphics[height=1.8in]{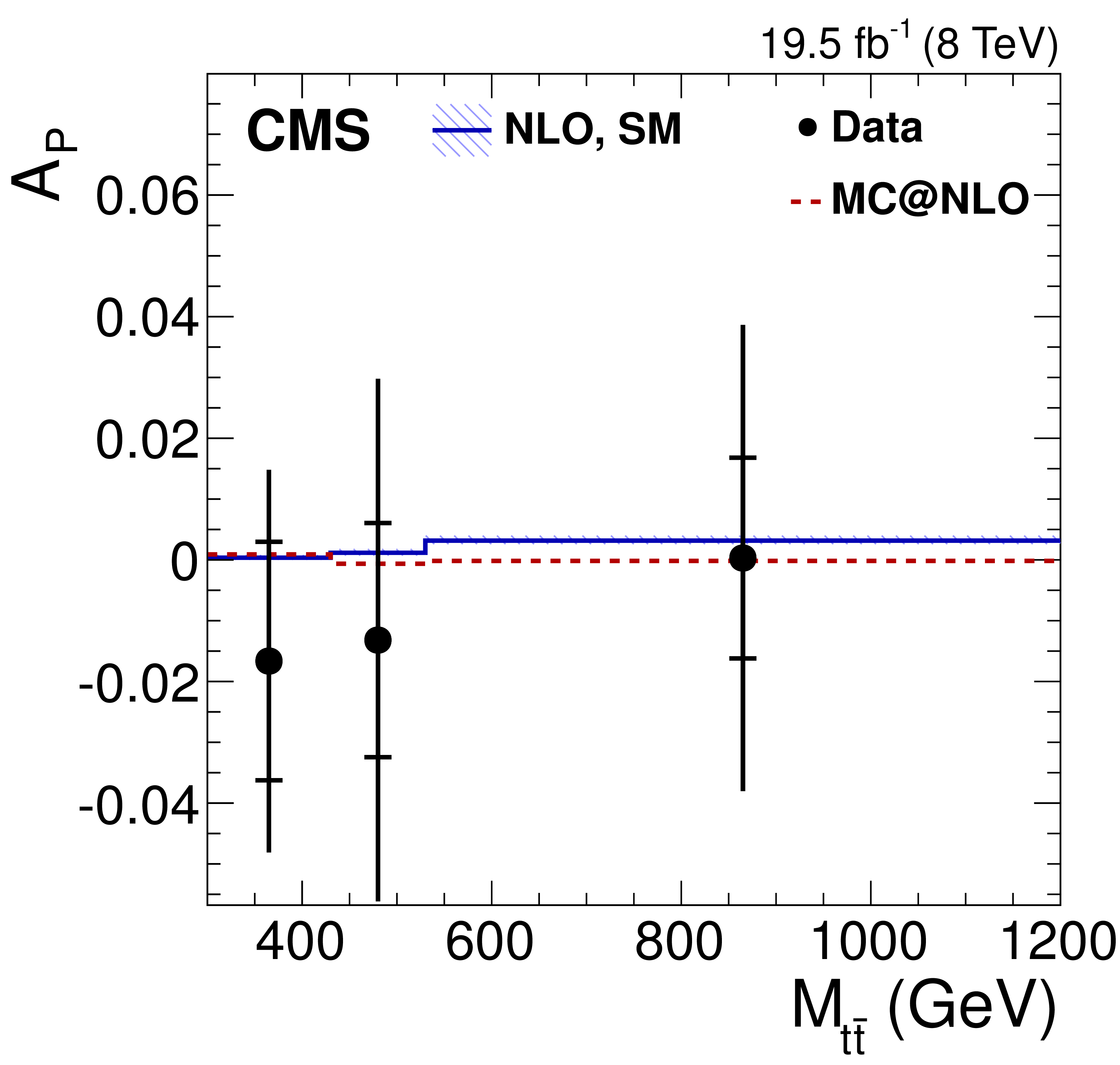}
\includegraphics[height=1.8in]{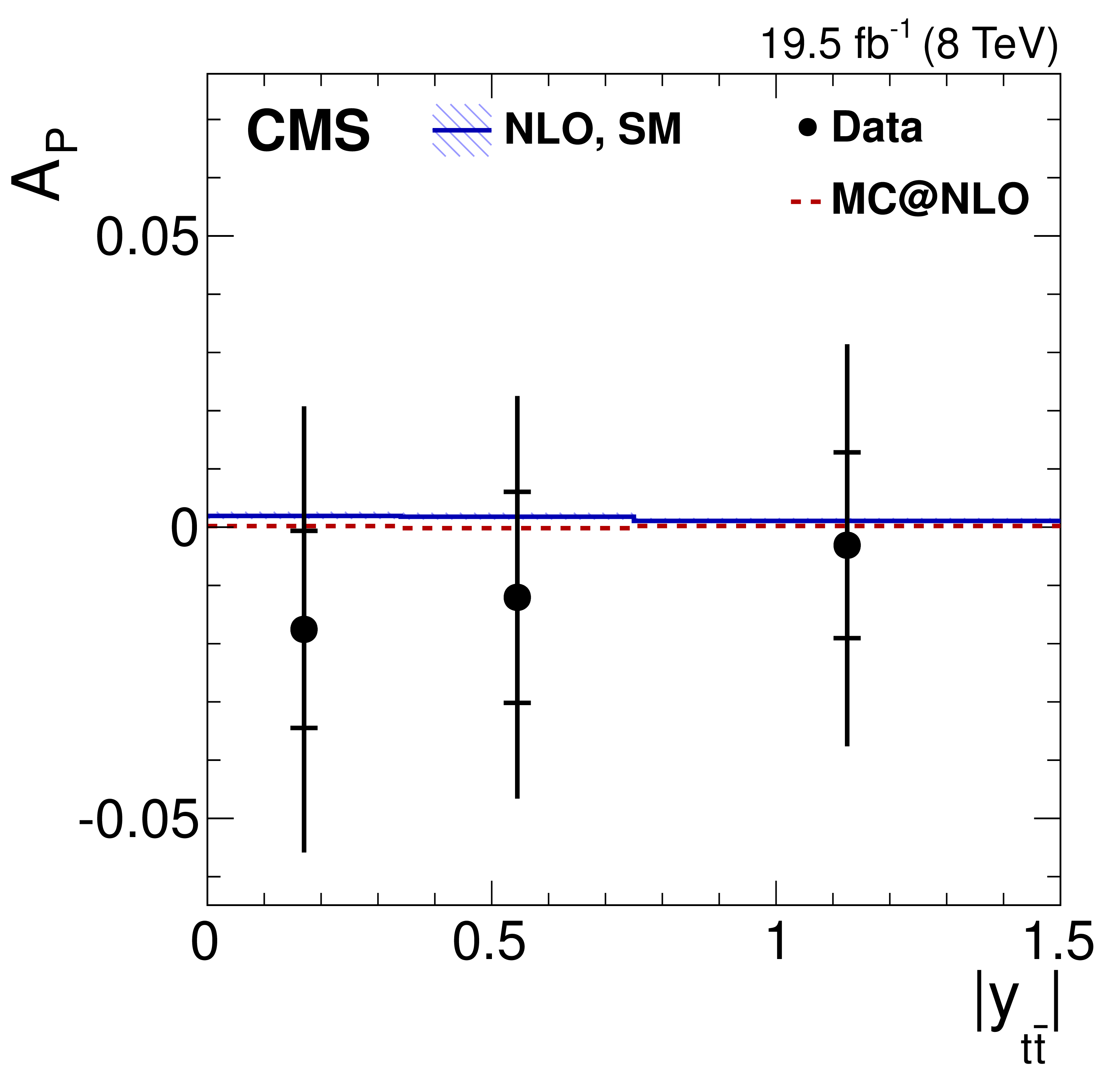}
\includegraphics[height=1.8in]{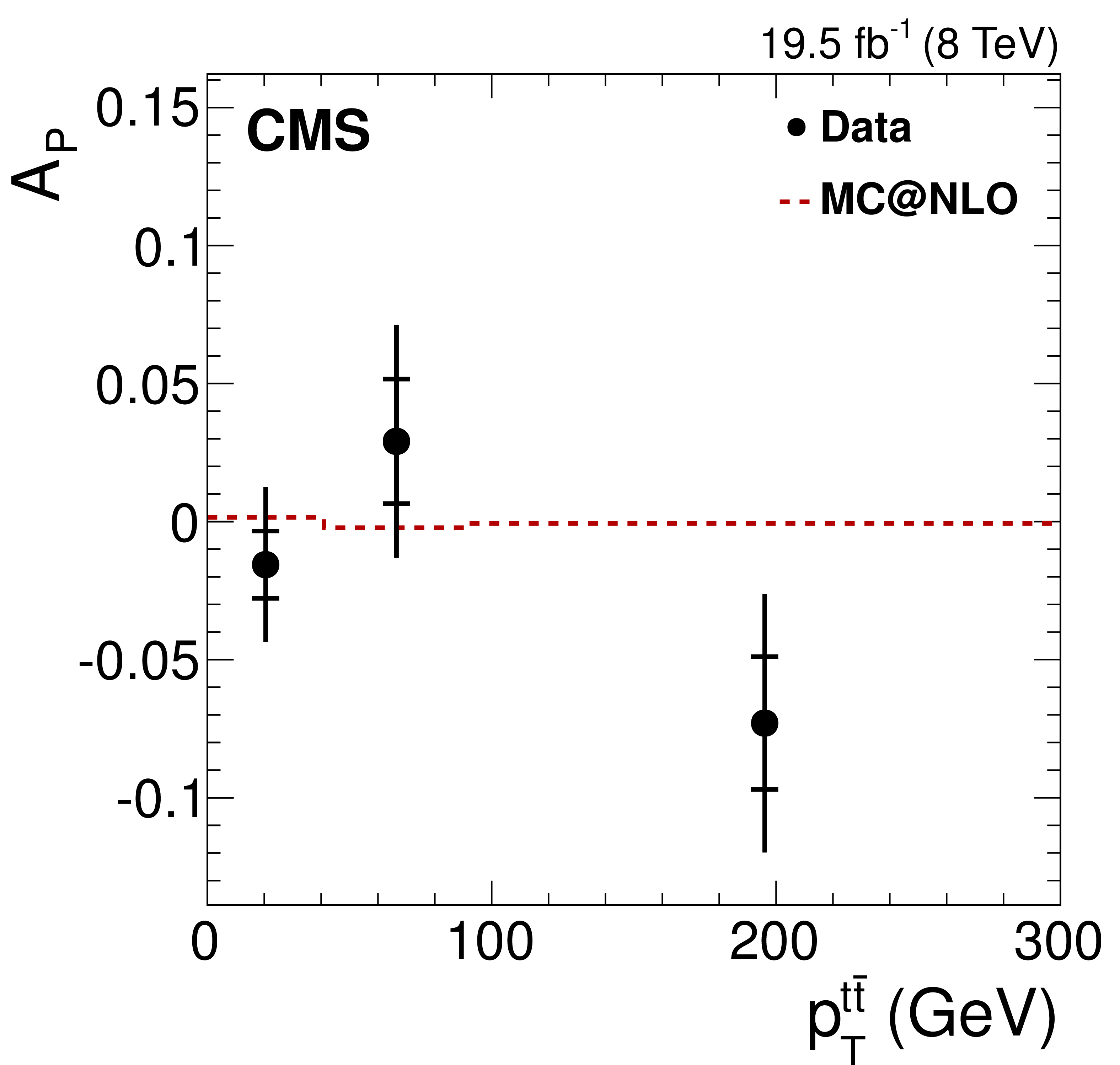}
\caption{Asymmetry variables, $A_{\Delta\phi}$ (1st row), $A_{cos\phi}$ (2nd row), $A_{c_1c_2}$ (3rd row), and $A_P$ (4th row), as a function of $M^{t\bar{t}}$ (left), $\eta^{t\bar{t}}$ (middle), and $p_T^{t\bar{t}}$(right). The unfolded double differential distributions (black dots) are compared to parton-level predictions from MC@NLO (dashed red lines), theoretical prediction at next to leading order with SM spin correlation (blue solid line), and similar theoretical prediction without the spin correlation (dotted blue). The inner (outer) bars in the data points represent statistical (total) uncertainties in the measurement. The hatched area in the theoretical calculation represent scale uncertainty from the systematic variation of $\mu_R$ and $\mu_F$ by a factor of two\cite{8TeVspincorr}.}
\label{fig:2Ddist}
\end{center}
\end{figure}

The final measurements were made at parton level by unfolding background subtracted distributions. Regularized unfolding method as implemented in TUNFOLD\cite{tunfold} was used to correct for the detector inefficiencies and acceptances. The unfolded one dimensional differential cross-section, as a function of $\Delta\phi_{l^+ l^-}$, $cos\phi$, $\cos\theta_{l^+}^*\cos\theta_{l^-}^*$, and $\cos\theta_{l}^*$, are provided in Figure~\ref{fig:1Ddist}. Each distribution in the figure is normalized to unit area and was found to be in agreement with the SM prediction. Using a similar procedure, two dimensional measurements of the asymmetries related to each of the spin correlation and the polarization quantities were also measured as a function of the pseudorapidity ($\eta^{t\bar t}$), transverse momentum ($p_T^{t\bar t}$), and mass ($M^{t\bar t}$) of the $t\bar t$ pair, and are shown in Figure~\ref{fig:2Ddist}. The inclusive values of these asymmetries were used to obtain spin correlation and polarization coefficients, $C_{hel}=0.278 \pm 0.084$, $D=0.205\pm0.031$, $P^{CPV}=0.000 \pm 0.016$, and $P=-0.022\pm0.058$ respectively. The systematic uncertainties in the measurements came from various experimental sources, modeling of the $t\bar{t}$ system, and unfolding. The major contributions to the experimental uncertainty in the asymmetry variables originate from the jet energy scale, the jet energy resolution, and the lepton energy scale. Top quark $p_T$ modeling, parton distribution functions, factorization and renormalization scales, top quark mass and hadronization contributed to the $t\bar{t}$ modeling uncertainty. The unfolding procedure also introduced some uncertainty; however, the regularization is found to have a minimal contribution to the total uncertainty. \\

\begin{figure}[htb]
\begin{center}
\includegraphics[height=2.8in]{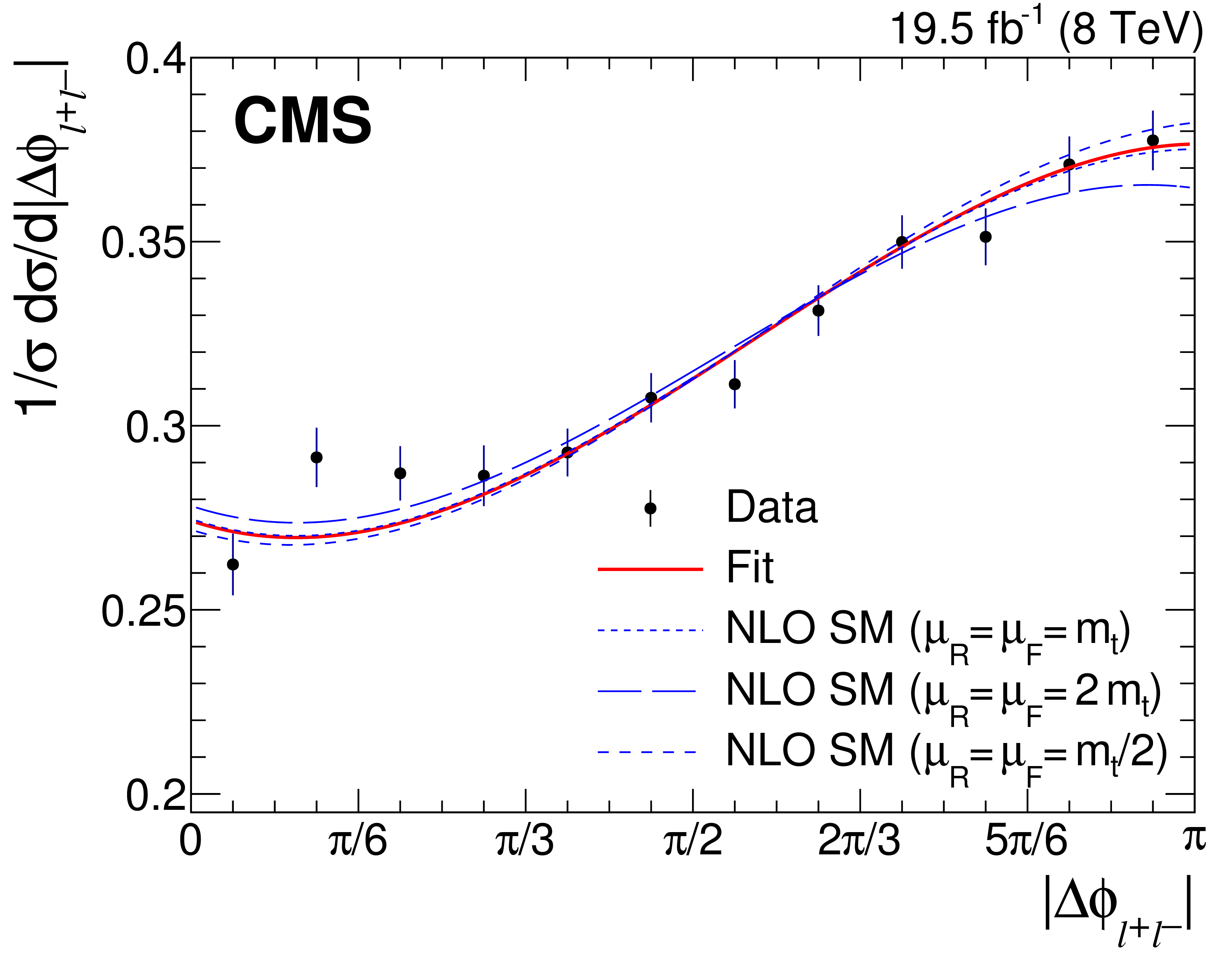}
\caption{Normalized $\Delta\phi_{l^+ l^-}$ differential cross section distribution from data (black dots), and the fit (red line) made from composite template of the SM distribution and NP contribution. The dashed lines show the scale uncertainty in the SM NLO prediction from the systematic variation of $\mu_R$ and $\mu_F$ by a factor of two\cite{8TeVspincorr}.}
\label{fig:limitplot}
\end{center}
\end{figure}

\noindent
Along with the SM validation, a search for hypothetical top quark anomalous coupling was done by performing a parametric template fit of the spin correlation distribution using a NP template with a SM template. The spin correlation distribution used was a projection of the $\Delta\phi_{l^+ l^-}$ distribution in the bins of $M^{t\bar t}$, in order to minimize systematic uncertainty. No evidence of new physics was observed as shown in Figure~\ref{fig:limitplot}. As a result, exclusion limits on the real part of the chromo magnetic dipole moment, Re($\mu_t$), and imaginary part of the chromo electric dipole moment, Im($d_t$), models were evaluated. At a 95\% confidence level, values outside of $-0.053<$ Re $(\mu_t)<0.026$ and $-0.068<$ Im $(d_t)<0.067$ were excluded. 

\section{Summary and Future Plans}
CMS performed top quark polarization and $t\bar{t}$ spin correlation measurement at $\sqrt{s}=8$ TeV using 19.5fb$^{-1}$ of data. Four observables were selected for the measurement. Associated asymmetries with respect to $M^{t\bar{t}}$, $\eta^{t\bar{t}}$, and $p_T^{t\bar{t}}$ were also measured. The measured differential distributions, associated asymmetries as well as the polarization and spin correlation values were found to be consistent with the SM prediction. At $\sqrt{s}=13$ TeV, there is an availability of $\sim$35fb$^{-1}$ of data. The ongoing analysis at the higher center of mass energy will measure all the independent coefficients of the spin dependent part of the $t\bar{t}$ production spin density matrix. These coefficients can be classified with respect to parity P, charge parity CP, naive time reversal $T_N$, and Bose symmetry\cite{bernreuther}; hence, providing full information on the underlying physics of $t\bar{t}$ production and decay. Furthermore, with approximately twice the amount of data, the ongoing analysis will make double differential cross section measurements of the variables with respect to quantities such at $p_T^{t}$, $E_T^{miss}$, $M^{t\bar{t}}$, which will allow for a better handling of the systematics, and also for probing new phase spaces that were not probed in earlier measurements.\\
\\
The CMS measurement at 8 TeV also searched for new physics contributions due to anomalous top coupling. Exclusion limits on the real part of the chromo magnetic dipole moment and imaginary part of the chromo electric dipole moment were evaluated. The ongoing CMS analysis plans to use a similar method of template fitting to constrain NP models including a supersymmetric extension of SM (SUSY) where stops are produced by a similar mechanism as tops are in the SM. Stops decaying via either the top quark channel $(\tilde{t}\rightarrow t\tilde{\chi^0}\rightarrow(W^+b)\tilde{\chi^0})$
or chargino channel $(\tilde{t}\rightarrow b\tilde{\chi^+}\rightarrow b(W^+\tilde{\chi^0}))$
will increase the measured $t\bar{t}$ cross section. However, in contrast to the SM scenario, the SUSY tops are expected to have non-zero polarization and uncorrelated $t\bar{t}$ pairs. The likelihood fit of the spin correlation and polarization distributions using templates from SUSY(or NP) and SM will, therefore, be used to constrain beyond the standard model.


%
%
%
%
%
 
\end{document}